\documentclass[prl,twocolumn, 10pt, reprint,nofootinbib]{revtex4-1}
\usepackage{graphicx}
\usepackage{epsf}
\begin{document}

\title{Dark matter seeding and the kinematics and rotation of neutron stars}

\author{M. \'Angeles P\'erez-Garc\'ia$^1$~\footnote{mperezga@usal.es} and  Joseph Silk$^2$~\footnote{j.silk1@physics.ox.ac.uk} }

\affiliation{$^1$ Departamento de F\'{i}sica Fundamental and IUFFyM, \\Universidad de Salamanca, 
Plaza de la Merced s/n 37008 Salamanca\\
$^2$ Institut d'Astrophysique, UPMC, 98 bis Boulevard Arago, Paris 75014,  France; Department of Physics and Astronomy, The Johns Hopkins University, Homewood Campus, Baltimore MD 21218, USA;  Beecroft Institute for Particle Astrophysics and Cosmology,  University of Oxford, Keble Road,  Oxford OX1 3RH, UK}

\date{\today}

\begin{abstract}
Self-annihilation of dark matter accreted from the galactic halo in the inner regions of neutron stars may affect their kinematical properties, namely velocity kicks and rotation patterns. We find that if a stable long-lived single or multiple strangelet off-center seed forms, there is an associated change in momentum and torque that may affect the kinematical observables of the star.
\end{abstract}
\maketitle
Light DM  particle candidates in the range of $4.5-12$ $\rm GeV/c^2$ are favoured by recent results in DM direct detection, including  the DAMA/LIBRA,  CoGENT and CRESST-2 experiments \cite{lightdm}, although these results have yet to be confirmed.
Some  DM  indirect detection experiments \cite{indirectdm} 
support these  indications. 
There is however  tension with other results, notably  Fermi dwarf limits  \cite{fermidwarf} and CMB damping tail measurements \cite{cmb}, but these can be avoided if for example there is an asymmetric DM component \cite{drees}.
%
In addition,  these particles  interact gravitationally and can be accreted onto stars. Here we explore some implications for neutron stars (NSs) that extend our earlier work  \cite{perez11}.
%

By studying the physics of  NSs, we find that macroscopic measurable quantities such as linear velocities and rotation periods  could provide an indirect indication of the presence and nature of DM.
 Typically, the average spatial velocity of a NS progenitor is less than $15$ $\rm km\, s^{-1}$, which means that they must receive a substantial {\it kick} at birth. Pulsars (PSRs) show very regular electromagnetic emission due to the misalignement of rotation and magnetic axes that make them visible to the intersecting light-of-sight observer on Earth. Surface magnetic fields for these objects are believed to be in the range $B \approx 10^9-10^{15}$ G. As  is well known, most  radio PSRs exhibit transverse velocities ranging from a few $100$ $\rm km$ $\rm s^{-1}$ to above $1000$ $\rm km$ $s^{-1}$, mostly in the galactic plane \cite{hobbs}. As for rotation, the observed PSR periods range from less than $0.1$ s to the {\it death line} at $\approx 1$ s. A different origin accounts for the recycled millisecond PSRs or magnetars with somewhat larger periods \cite{woods}. It has been shown in recent simulations \cite{spruit} that rotation and kicks at birth can be linked to the detailed mechanisms in supernova (SN) core collapse and this,  unless the kick is head-on, may cause a change in the rotation pattern. One can obtain modest kicks in the $100-500$ $\rm km$ $\rm s^{-1}$ range \cite{3dkick} based on an anisotropic momentum distribution of the ejecta or parity-violating scattering of neutrinos \cite{pv-neu}.

Previously \cite{perez11}, we showed that the effect of DM seeding in NSs  may provide a mechanism to form strange quark matter (SQM) bubbles or strangelets that, if stable and long-lived, could induce a  partial NS conversion into a hybrid SQM star. This effect is somewhat analogous to what happens in nuclear ignition models where sparks can be induced in the bulk by heating a nuclear plasma with a laser \cite{fusion}.  Terrestrial strangelet production searches have been performed at high energy colliders at E864 in BNL and are currently undertaken at LHC and ISS (International Space Station) \cite{search}. We show here  that DM single or multispot spark seeding in compact objects  may lead to a modification of NS kinematics and rotation. A typical NS has a mass $M_{NS}\approx 1.4 M_{\odot}$, radius $R_{NS}\approx 10\,\rm km$ and velocity, $v_{NS}$, that can be in excess of $\approx 1000$ km/s. The associated momentum $p_{NS} \approx M_{NS} \, v_{NS} \approx 3 \,10^{33} \rm g \,10^{8}\, \rm cm/s\approx 3 \, 10^{41}$ g cm/s. The birth energy released in a SN type II event is known to be  $E_{\nu}\approx 10^{53}\, \rm erg$. If kicks were to be explained by (massless) neutrinos, this means that they should carry typically $p_{\nu} \approx E_{\nu}/c \approx 10^{53}/3\, 10^{10} = 3.3\, 10^{42}$ g cm/s. This results in an asymmetry  $\alpha_p=p_{NS}/p_{\nu} \approx 10^{-2}$. However, this neutrino {\it rocket} mechanism may be residual since thermal effects tend to make emission isotropic. Recent analyses of PSR observations seem to indicate possible correlations for the directions of the spin axis and the velocity vector,  another piece of evidence for a large velocity in 3D phase space \cite{align}. The radial distribution of the birth location of NSs at given galactocentric distance $R$ can be fit using a relation derived by Lorimer that includes selection effects \cite{lorimer}
\begin{equation}
\rho_{PSR}(R)=A \left(\frac{R}{\rm kpc}\right)^{n}e^{-(\frac{R}{R_0})} \,\rm kpc^{-2},
\end{equation}
where $n=2.35$, $R_0= 1.528\,\rm kpc$ and $A = 64.6$ $\rm kpc^{-4.35}$ and  peaks at $R_{max}=3.6$ kpc. NSs can accrete DM from the galactic halo by gravitational capture. This DM profile can be characterized using the continuously-varying slope function \cite{profile},
\begin{equation}
\rho_{DM}(r)=\rho_{-2}e^{\frac{-2}{\alpha} [ (\frac{r}{r_{-2}})^{\alpha} -1]},
\end{equation}
with $\rho_{-2}=0.22 \,\rm GeV/cm^3$, $\alpha=0.19$, and $r_{-2}=16 \,\rm kpc$ so that at the solar neighbourhood the Keplerian velocity is $v \approx 220$ km/s and the local DM density is $0.7\, \rm GeV/cm^3$. Both observables can be  approximately fit assuming most of the matter is DM and using  typical parameters of Milky way-size halos. DM  is accumulated over time in the central NS regions where it is rapidly thermalized ($t_{th}\approx 0.13 \, \mu s$ to achieve  $T=30$ MeV or $t_{th}\approx 4 \, \mu s$ to achieve  $T=1 \,\rm MeV$ \cite{nussinov} in the hot or cooled NS, respectively). Assuming DM is well described by a Boltzmann distribution, the thermal radius is given by 
\begin{equation}
r_{th}(t)=\left( \frac{3 k_B T_{c}(t)}{2 \pi G \rho_c(t) m_{X}}\right)^{1/2},
\end{equation}
where $m_X$ is the DM particle mass. This quantity is time-dependent since the central temperatures $T_c$ and mass densities $\rho_c$ evolve with those in the  progenitor star. For example, during the pre-collapse, typically $T_{c}=0.001-1\,\rm MeV$ and $\rho_{c}=10^{10-13}$ $\rm g/cm^3$ while in the bounce phase, the temperature increases to $T \approx 40$ MeV and there is a dramatic increase in the central density up to $\rho_{c}=4\, 10^{14}$ $\rm g/cm^3$ in the  time scale of the gravitational collapse, $t_{collapse}\approx 10^{-3}$ s. Assuming DM is of the Majorana type, it may self-annihilate. We define $\alpha_r=\frac{r_{th}}{R_{NS}}$ as a measure of the radial coordinate fraction involved in annihilation processes that, at the few GeV scale, involve photons, leptons , light $q{\bar q}$ pairs \cite{kuhlen}. The ratio of  the thermalization volume to that in the NS, $V_{th}/V_{NS}\approx \alpha_r^{3} \approx 10^{-6}$. The central annihilation or {\it spark} production rate at radius $r$ will be given by $\frac{d\Gamma_{\rm a}(r)}{dr} \approx \left <\sigma_{a} v\right> n_X^2 4 \pi r^2$ where we assume that the DM particle number density, $n_{X}(r)\approx n_{X} $, is constant on a scale $r \approx 10^2$ m. In a SN type II core collapse, peak temperatures and densities are achieved off-center and production of one or more sparks may subsequently happen in those regions. Other exotic mechanisms may induce thermal or quantum nucleation of quark bubbles \cite{exotic}, however, it turns out that this may not be efficient since typical timescales are larger than standard cooling scenarios. The energy released  over the thermalization volume and, accordingly, the rate of spark formation  can be  obtained assuming an efficiency, $f$, for DM annihilation \cite{perez11} as ${\dot E_{\rm a}} \approx 2 f{\cal C} m_{X}c^2$ where ${\cal C}$ is the DM capture rate. Assuming NS matter is mostly in the core under deconfined nucleons, the capture rate  can be written as \cite{Gould},
 \begin{equation}
{\cal C}\approx \frac{2.7 \times 10^{29}}{ m_{X} (GeV)} \frac{\rho_{DM}}{\rho_{0}}\, (s^{-1}),
\label{caprate}
\end{equation}
where we have used  $\sigma_{XN}=7\,10^{-41}\,\rm cm^2$ and  average core density $\bar \rho_{NS}=\frac{M_{NS}}{V_{NS}}$. We consider the spatial NS distribution speak and the Keplerian velocity is calculated from the enclosed DM and baryonic (B) mass components as $v(R_{max})=\sqrt{\frac{G \, M_{DM+B}(R_{max})}{R_{max}}}$. Galaxy models \cite{klypin} show an inner mass distribution where $f_M=M_B/M_{DM}\approx 1$ , then at location $r$, $M_{DM+B}(r) \approx M_{DM}(r)+{\bar M_B}=\int_0^r 4 \pi r'^2 \rho_{DM}(r') dr' +{\bar M_B}$ where ${\bar M_B}=\int {\bar \rho_B} \frac{d\Omega}{4 \pi} $ This yields $v \approx 220$ km/s at the solar circle and at the NS distribution peak $v \approx 190$ km/s. The spark energy release $\Delta E_{spark}=2 f m_X c^2$ will be deposited in the medium in the vicinity of the production site. The primordial spark energy ratio to nucleon mass ($m_N\approx 1 \,\rm GeV$) is $\Delta E_s/m_N \approx 2 m_X  f/m_N \approx {\cal O}(1-10)$. This high energy deposition may potentially lead to the formation of a strangelet \cite{perez11}. The binding energy of such a strangelet is given in the MIT model as \cite{mit} $E^A_{\rm slet}\approx E^A(\mu_i, m_i, B)+E_{Coul}$ where $\mu_i$ and $m_i$ are the chemical potential and mass of the ith-type quark, respectively. B is the MIT bag constant and $E_{Coul}$ is the correction due to electrical charge. Those with mass number $A>A_{min}$ are long-lived and energetically stable. As an estimate, the production rate of long-lived strangelets (with $A=10$) at $R_{max} $ would be ${\dot N_{\rm slet}}={\dot E_{a}}/E^A_{\rm slet}\approx 10^{\rm 27}  \,s^{-1}$ for a density of about $2 n_{0, A}$ ($n_{0, A}=0.17\, \rm fm^{-3}$ or equivalently mass density $\rho_0=2\,10^{14}\, \rm g/cm^3$). The ratio of spark energy to  strangelet binding energy $BE=E^A_{\rm slet}-Am_N$ is $\Delta E_s/BE \approx {\cal O}(1-10)$ making strangelets energetically accessible systems.

We will consider now for simplicity a ${\it single}$ spark seed model located off-center at ${\bf r_{off}}=r_{off}\,{\bf e_{off}}$, $r_{off} \leq \alpha_r R_{NS}$.  From the typical conditions in the interior of the (proto-)NS  in the location of the growing spark $\alpha_r \approx 10^{-2}-10^{-1}$. The spark radial  direction forms an angle $\beta$ with respect to ${\bf {e}_{spin}}$, unitary vector in the spin axis direction. Once formed, the quark bubble is composed of burnt material consisting of u, d and s light quark sectors. The seed radius $R_s$ will grow as a function of time, $t$, and central distance, $r$, due to the strong and weak interactions at a rate $dR_{s}/dt=v_b(t, r)$ that can be estimated assuming a simplified ${\it two-step}$  burning front that could  evolve dynamically \cite{bhatta} depending on the matter equation of state (EOS). Since the strong interaction time-scale $t_{strong} \le {\cal C}^{-1}$,  the annihilation mechanism may produce an effective local heat source.  Assuming the minimum stable strangelet has baryon number $A_{min}$, then in a semiclassical approach there is an increase in the energy release per nucleon, $\Delta e_{th} \approx \frac{\Delta E_s}{A_{min}} \approx \frac{2 f m_X c^2}{A_{min}} \approx 0.1-1 \, \rm GeV \approx {\cal O}(\Lambda_{QCD})$ which is of the order of the binding energy of the nucleon or QCD scale, $\Lambda_{QCD} \approx 0. 217$ GeV. 

In order to obtain the effect on the kinematic NS observables, we  consider the combustion of an electrically charge neutral system of hadronic unburnt matter  (phase 1) into the burnt quark phase (phase 2). For the hadronic star (composed of neutrons, protons and electrons in beta equilibrium and in global electrical charge neutrality), the EOS $P=P(\epsilon, \rho)$ will be taken as the APR model \cite{APR} at $T\approx 0$. No magnetic fields will be considered. Instead, for the deconfined quark burnt material the MIT bag model will be taken for consistency with the finite lump of quark matter description. As the combustion takes place, weak interactions act on a time-scale of $t_{weak}\approx 1/\Gamma_{Kaon} \approx 10^{-8}$ s and a $u-d-s-e$ system forms while the gravitational force  tends to contract the star on a timescale $t_{grav}\approx 2 R_{NS}/c=10^{-4}$ s. Non-zero strangeness fraction  involves reactions including $u+d\rightarrow u+s$, $u+e\rightarrow d+ \nu_e$, $u+e\rightarrow s+\nu_e$, where neutrinos freely escape. Jump conditions near the front in this relativistic fluid are given by  a set of conservation laws (energy-momentum and  baryon  current ) across the combustion front and the increase in entropy condition for the two phases. In the rest frame for the burning front ($c=1$) \cite{landau87} the set is written as $(\epsilon_1+p_1) v_1^2+p_1=(\epsilon_2+p_2) v_2^2 +p_2$ , $
(\epsilon_1+p_1) v_1=(\epsilon_2+p_2) v_2$, $s_1 v_1 \le s_2 v_2$, $n_1 v_1=n_2 v_2$, where  $p_i, \epsilon_i, n_i=\rho_i/m_i$ and $v_i$ are the pressure, energy, number density and velocity for the ith-phase. Using the preceding equations results  in $v_1^2=\frac{(p_2-p_1)(\epsilon_2+p_1)}{(\epsilon_2-\epsilon_1)(\epsilon_1+p_2)}$ and $v_2^2=\frac{(p_2-p_1)(\epsilon_1+p_2)}{(\epsilon_2-\epsilon_1)(\epsilon_2+p_1)}$. It is possible to classify the conversion when $v_1, v_2$ are comparable to the sound velocities for the two phases, $c^2_{s, i}=\frac{\partial p_i}{\partial \epsilon_i}$. According to accurate calculations \cite{madsenPRL}, finite strangelets may be positively charged on their surface due to the strangeness content in their deep interior for $A \approx 10^2-10^8$ and cannnot bridge the gap to large A systems. This may indicate a limited  growth potential in the context of safe finite volume ultrarelativistic heavy ion collisions in vacuum, however, in the core of these astrophysically-sized scenarios, the pressure is very large and hybrid objects may exist.
\begin{figure}[hbtp]
\begin{center}
\includegraphics [angle=-90,scale=0.65] {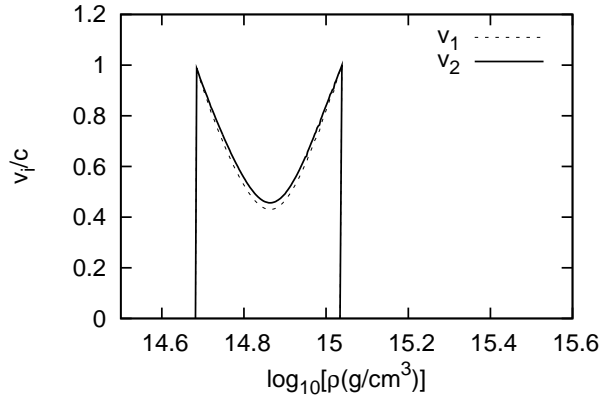}
\caption{Hadronic, $v_1$, and quark matter, $v_2$, burning velocities as a function of matter mass density. The burning front has a velocity $v_b=-v_1$.}
\label{Fig1}
\end{center}
\end{figure}
The radial velocity profile in the star can be constructed by solving the static Tolman equations \cite{tov} or their rotating counterparts. We will not consider the more realistic rotating profile and, instead, we will use the $r$-dependent velocity in the  burning site as in \cite{bhatta} as $v(r)=\int_{r_{off}}^r \frac{dv}{dr'} dr'$. Once formed, the growth of the quark lump can build up a net excess of momentum $\Delta {\bf P}$ in the system as a result of the different pressures in the burnt and unburnt regions, ${\Delta P_{12}}=P_1-P_2$. The associated impulse, ${{\bf \cal{I}}}$, and force, ${\Delta \bf F}$ relate as ${\bf \cal I}=\Delta {\bf F} \Delta t=\Delta {\bf P}$ with $\Delta t$ the time over which the thrust is acting. We make the simplifying assumption that the burning front remains spherical and moves  outwards at $v_b(r)=-v_1(r)$ and reaches the edge $R$ at $\Delta t=\int_{r_{off}}^R \frac{dr}{v_b(r)}$ in the comoving frame. Due to the fact that  the lump is evolving in a non-inertial frame there will be associated forces and, for simplicity, we adopt  the weak field Newtonian approximation. The mechanical force acting in the radial direction can now be parametrized using the effective transverse area $\Delta S=\pi R^2_s(t)$ as $\Delta {\bf F}_{mech}=\Delta P_{12}{\Delta S} {\bf e_{off}}$. There will be a gravitational acceleration component due to the unburnt mass $M(r)$ acting on the quark lump with volume $V_s=\frac{4}{3} \pi R^3_s$ and mass $M_{s}(R_s)=V_s(R_s) \rho_2(r_{off})$ and therefore $\Delta {\bf F}_{grav}= M_s(R_s) {\bf g}$ where ${\bf g}=- \frac{G  M(r_{off})}{r_{off}^2}{\bf e}_{off}$, using the fact that the unburnt progenitor star component is  $M(r_{off})=V(r_{off}) \rho_1(r_{off})$.  Due to the fact the compact star is rotating with angular velocity $\bf \Omega$, there is an additional effect due to centrifugal force $\Delta {\bf F}_{centrif}$ as the quark lump grows towards the star edge. This can be estimated as $\Delta {\bf F}_{centrif}=-M_s(R_s) \,{\bf \Omega} \times {\bf \Omega} \times {\bf r_{off}}$. Finally, $\Delta {F}_{centrif}=V_s(R_s) \, \rho_2(r_{off}) \, \Omega^2 r_{off} \, Sin \beta$.
Additionally there will be a contribution from the Coriolis force $\Delta {\bf F}_{cor}$ due to the motion of the burning front that will induce some extra rotation. The form adopted is $\Delta {\bf F}_{cor}=-2 M_s (R_s){\bf \Omega} \times {\bf v_b}$. That is $\Delta { F_{cor}}=V_s(R_s)\, \rho_2(r_{off}) \,{\Omega} \, v_1(r_{off})$. For the sake of simplicity we will now consider motion in the equatorial plane and set $\beta=\pi/2$. Then, the composition $\Delta {\bf F}=\Delta {\bf F}_{mech}+\Delta {\bf F}_{grav}+\Delta {\bf F}_{centrif}$ will produce an impulse that can impart a velocity kick in the progenitor NS with mass $M_{NS}$ as,\begin{equation}
\Delta \bf v_{NS}=\frac { \bf \cal I}{M_{NS} }=\frac{\Delta {\bf P}} {M_{NS} }.
\end{equation}
 Then the combustion will  partially (or fully) exhaust the remaining unburnt material and then $\Delta {\bf F} \rightarrow 0$. Assuming that in the final configuration, the object has a moment of inertia $I_{f}$ the angular velocity is affected in the conversion, since there must be a torque ${{\bf \tau}}$ modifying the angular momentum ${\Delta {\bf J}}={\bf \tau}{\Delta t}={{\bf r_{off}}} \times {\Delta \bf P}$. Using $\Delta {\bf F}=\Delta {\bf F}_{cor}$ these expressions we finally derive,
\begin{equation}
\frac{\Delta \Omega}{{\Omega}}=\frac{\Delta \tau \Delta t}{I_{f}{ \Omega}}.
\end{equation}
The moment of inertia $I_{f}$ will be, in general, time-dependent and smaller than the original due to EOS softening in the conversion. As an approximate parametrization to this value, we take $I_{f}=d M_{NS} R^2_{NS}$ with a $d\approx 0.18$. It is important to note that if the lump is not able to grow within the star structure due to energetics then the star configuration could be hybrid \cite{glendenning}. This event could provide an efficient source of gravitational waves and set a wave-form pattern with clear signatures of the conversion. The possible contribution to the jet-like formation in the initial stages of a NS could be somewhat influenced by internal shocks and crust ejection if the conversion is complete \cite{daigne-perez-silk-jet}.

\begin{figure}[hbtp]
\begin{center}
\includegraphics [angle=0,scale=0.8] {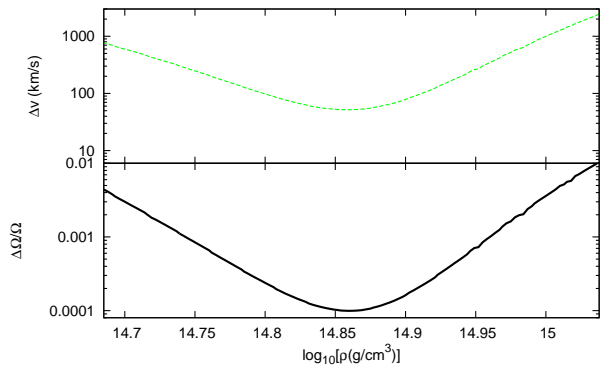}
\caption{Average velocity kick in km/s (upper panel) and relative change in angular velocity (lower panel) as a function of hadronic mass density, $\rho$.}
\label{Fig2}
\end{center}
\end{figure}
In Fig. \ref{Fig1} we can see the allowed physical velocities of the two phases as obtained for matter in beta equilibrium using the hadronic APR EOS \cite{APR} and the MIT bag model  EOS for the quark phase as a function of the logarithm of the  mass density. We use an intermediate value for the bag constant range, $B^{1/4}=150 \,\rm MeV$, and $T=0$, where long-lived strangelets can form. The solutions are constrained to the physical values $0 \le v_i^2\le 1$. It can be seen that for densities $\rm log_{10}\, [\rho(g/cm^3)] \le 14.685$ burning is not allowed. The particular density window could be EOS-dependent but we expect this feature to be a general trend. 
In Fig. \ref{Fig2} the variations in linear velocities $\Delta v$ in km/s (upper panel, dashed line) and angular velocity ratio $ \Delta \Omega/\Omega$ (lower pannel, solid line) are shown as a function of mass density. We have assumed a typical initial value for frequency of $\nu=300$ Hz. We  see that the linear velocities obtained are indeed bimodal, in the sense that both high kicks and low kicks are possible and are correlated with the change in the angular velocity. These kicks are compatible with NS observed values and also with measurements of the change in the angular velocity during GRB emission where Newtonian estimations give  $\Delta \Omega/\Omega \approx 10^{-3}-10^{-2}$ \cite{rot1}. We have used on average $t_{thrust} \approx 0.3$ ms according to radial profile estimates of $\approx 1$ km burning propagation in the allowed density window.
We have shown that if a stable quark seed is able to grow, due to the annihilation of DM particles in the central regions of compact objects and according to the EOS of matter, then a kinematical signature should be present giving  both bimodal low and high kicks and spin angular velocity values.
We would like to thank J. Miller and C. Albertus for helpful discussions. M. A. P. G.  would like to thank University of Oxford for its kind hospitality. She acknowledges partial support under ESF project COMPSTAR, MICINN projects Consolider MULTIDARK, FIS-2009-07238 and Junta de Castilla y Le\'on GR-234. 
\section{bibliography}

\end{document}